\documentclass[aps,prl,reprint,nofootinbib,superscriptaddress]{revtex4-1}

\usepackage[utf8]{inputenc}
\usepackage{color}
\usepackage{amsmath,amssymb}
\usepackage{amsfonts}
\usepackage{verbatim}
\usepackage{makeidx}
\usepackage{slashed}
\usepackage[vcentermath]{youngtab}
\usepackage{float}
\usepackage{booktabs}
\usepackage{graphicx}
\usepackage{mathrsfs}
\usepackage{bm}
\usepackage{braket}
\usepackage{url}
\usepackage{etoolbox}
\apptocmd{\sloppy}{\hbadness 10000\relax}{}{}
\allowdisplaybreaks[3]
\usepackage{color}

\newcommand{\be}{\begin{eqnarray}}
\newcommand{\ee}{\end{eqnarray}}

\usepackage{ulem}
\definecolor{ao}{rgb}{0.0, 0.5, 0.0}

\definecolor{MO}{rgb}{0.0, 0.0, 0.8}

\begin{document}

\title{Doubly Heavy Tetraquark Resonant States}

\author{Q.~Meng}
\affiliation{Department of Physics, Nanjing University, Nanjing 210093, P.R.~China}
\author{M.~Harada}
\affiliation{Department of Physics, Nagoya University, Nagoya, 464-8602, Japan}
\affiliation{Kobayashi-Maskawa Institute for the Origin of Particles and the Universe, Nagoya University, Nagoya 464-8602, Japan}
\affiliation{Advanced Science Research Center, Japan Atomic Energy Agency, Tokai 319-1195, Japan}
\author{E.~Hiyama}
\affiliation{Department of Physics, Tohoku University, Sendai 980-8578, Japan}
\affiliation{Nishina Center for Accelerator-Based Science, RIKEN, Wako 351-0198, Japan}
\affiliation{Advanced Science Research Center, Japan Atomic Energy Agency, Tokai 319-1195, Japan}
\affiliation{Research Center for Nuclear Physics (RCNP), Osaka University, Ibaraki 567-0047, Japan}
\author{A.~Hosaka}
\affiliation{Research Center for Nuclear Physics (RCNP), Osaka University, Ibaraki 567-0047, Japan}
\affiliation{Advanced Science Research Center, Japan Atomic Energy Agency, Tokai 319-1195, Japan}
\affiliation{Nishina Center for Accelerator-Based Science, RIKEN, Wako 351-0198, Japan}
\author{M.~Oka}
\affiliation{Advanced Science Research Center, Japan Atomic Energy Agency, Tokai 319-1195, Japan}
\affiliation{Nishina Center for Accelerator-Based Science, RIKEN, Wako 351-0198, Japan}



\begin{abstract}
Spectrum of the doubly heavy tetraquarks, $bb\bar q\bar q$, is studied in a constituent quark model.
Four-body problem is solved in a variational method where the real scaling technique is used to identify 
resonant states above the fall-apart decay thresholds.
In addition to the two bound states that were reported in the previous study we have found 
several narrow resonant states above the $BB^*$ and $B^*B^*$ thresholds.
Their structures are studied and are interpreted by the quark dynamics. 
A narrow resonance with spin-parity $J^P=1^+$ is found to be a mixed state of a compact tetraquark  
and a $B^*B^*$ scattering state. This is driven by a strong color Coulombic attraction between the $bb$ quarks.
Negative-parity excited resonances with $J^P=0^-$, $1^-$ and $2^-$ form a triplet under the heavy-quark spin symmetry.
It turns out that they share a similar structure to the $\lambda$-mode of a singly heavy baryon as a result of the strongly attractive correlation for the doubly heavy diquark.  
\end{abstract}

\date{\today}
\maketitle

\section{Introduction}

Dynamical structure of multi-quark hadrons, made of more than three quarks, is one of the most interesting subjects in hadron physics. Due to the color confinement in quantum chromodynamics (QCD), hadrons composed of quarks (and gluons) must be ``white'', or color $SU(3)$ singlet. Conventional hadrons are made from a quark and an anti-quark (mesons), or three quarks (baryons), while many more color-singlet combinations of quarks are allowed for multi-quark 
systems.~\cite{GellMann:1964nj,Chen:2016qju,Hosaka:2016pey,Esposito:2016noz,Guo:2017jvc,Ali:2017jda,Yamaguchi:2019vea}

Since the epoch-making observation of $X(3872)$ by Belle collaboration in 2003~\cite{Choi:2003ue}, 
many candidates of new tetra-quark ($qq\bar q\bar q$) 
and penta-quark ($qqqq\bar q$) hadrons were reported~\cite{Aaij:2015tga,Aaij:2019vzc}. 
Such multi-quark hadrons must help us to resolve the mystery of color confinement dynamics, 
as they contain new types of color configurations appearing only 
in the systems of more than three quarks.

So far many observed candidates of multi-quark states are in a scattering region, in particular at the vicinity of thresholds of two or more particles.  
In such a situation a relevant question is  whether they are 
resonances or threshold effects\cite{Hanhart:2007yq,Braaten:2007dw}, 
or possibly if there are any stable bound states below the thresholds.   
The present paper reports the results of our attempt to answer these questions 
focusing on tetraquark resonances.  
This is the next to our former work for bound states.  

To start with, let us briefly review previous works relevant to the present discussions. 
In general, systems with heavy quarks have advantages to form bound states, 
as the kinetic energy is suppressed due to a large reduced mass. 
This expectation has already been confirmed in the quarkonium  spectra; 
bottomonia are more stable states than charmonia.
Lattice QCD also supports the idea so that a few bound states are predicted in multi-strange/charm dibaryons (6 quarks).\cite{Iritani:2018sra,Sasaki:2019qnh,Gongyo:2017fjb} 

For the tetraquark systems, several possible bound states 
have been studied.~\cite{Ader:1981db,Vijande:2009kj,Caramees:2018oue,Wang:2019rdo}
To realize stability, their masses should be lighter than 
the threshold energy of two (or more) hadrons to which the tetraquark states would decay. 
It turns out that stable states are realized for 
doubly-heavy tetraquarks, $QQ\bar q\bar q$~\cite{Karliner:2017qjm,Eichten:2017ffp,Quigg:2018eza}.
In contrast, hidden-heavy tetra-quarks of $Qq \bar Q\bar q$ are not stable.  
The possible decay channels for doubly- or hidden-heavy tetraquarks are two mesons of 
($Q\bar q$)-($Q\bar q$) or ($Q\bar Q$)-($q\bar q$), respectively.
Due to the strong color-Coulomb attraction between heavy quarks, 
twice larger for $Q\bar Q$ than for  $QQ$, 
the ground state of the hidden-heavy system is a system of heavy and light quakonia, 
($Q\bar Q$)-($q\bar q$), while 
the doubly-heavy tetraquark survive as a stable state without decaying into two heavy mesons ($Q\bar q$)-($Q\bar q$).  


Among many studies, we have investigated systematically tetraquarks 
with various quark combinations from light to heavy flavor regions, 
and found several stable bound states for doublely-heavy tetra-quark $QQ\bar q\bar q$ systems~\cite{Meng:2020knc}.  
In particular, two S-wave $bb\bar q\bar q$ bound states with the isospin and spin-parity $(I, J^P)=(0, 1^+)$ 
are found, one deeply bound and another a shallow bound states slightly below the lowest two-particle threshold ($B +B^*$).
The deep one is a compact tetra-quark bound state, 
while the shallow one is extended in space and consistent with a molecular bound state of $B(b\bar q, 0^-)$ and $B^*(b\bar q, 1^-)$.
It is remarkable to see that a single Hamiltonian gives both the compact and loose bound states of two different natures.
The obtained results have been compared to the available lattice results and verified that 
they are consistent 
with each other~\cite{Francis:2016hui,Junnarkar:2018twb,Leskovec:2019ioa,Hudspith:2020tdf,Mohanta:2020eed,Bicudo:2021qxj}.
This encourages us to go further the study to the investigations of resonances.  

In the present as well as in the previous studies, we employ the standard quark model Hamiltonian~\cite{SilvestreBrac:1996bg}
and diagonalize it by a full four-body method including couplings to fall-apart scattering channels for resonances.
A caveat is for the applicability of the present quark model 
which is quenched without $q\bar q$ creation or annihilation, or without meson couplings.  
Therefore, our discussions should be limited to the low energy region 
where meson, in particular, (multi) pion emission is suppressed.  


Now a novel result in the present study is that narrow $bb\bar q\bar q$ resonances are found, 
one is isospin $I=0$ and spin-parity $J^P = 1^+$, and 
the other isospin $I=0$ with three degenerate spin-parity states of $0^-, 1^-, 2^-$.  
The latter is the first observation of the heavy quark spin triplet in the heavy tetraquark systems.
Its existence in the pentaquark systems was pointed out in Ref.~\cite{Shimizu:2018ran}. 
Interestingly the positive parity one appears lower which is only about 50 MeV above the decaying channel of 
$BB^*$, and the negative parity one higher at 145 MeV above $BB^*$.  
These energies seem unexpectedly small.
They are, however, understood as the realization of the combination of two dynamics with different scales, 
one is the strong color-Coulomb attraction for heavy diquark $QQ$, 
and the other the strong interaction of order $\Lambda_{QCD}$ for light anti-diquark $\bar q \bar q$.  
Consequently, they develop a unique diquark clustering structure for both heavy and light quarks.  
In short, positive parity resonance develops
an internal nodal excitation of the $QQ$ diquark, while the negative parity one 
is a $\lambda$-mode like of the singly-heavy baryon, a quasi two-body mode of 
heavy diquark and light good diquark $(QQ)$-$(\bar q \bar q)$.  
Our results implies that the systematic study of masses and decay widths 
can access rich information for the internal structure of multi-quark systems.

%


\section {Formulation}

Following the method in our previous work~\cite{Meng:2020knc},
we employ a standard quark model Hamiltonian and solve the four-body problem
of two heavy ($Q=b$) quarks and two light anti-quarks ($\bar q = \bar u$ or $\bar d$).  

The Hamiltonian is given by
\begin{eqnarray}
\label{Hamiltonian}
\begin{aligned}
	H=&\sum_{i}^{4}\Big(m_i+\frac{{\boldsymbol{p}_i}^{2}}{2m_i}\Big)-T_G \\
	&-\frac{3}{16}\sum_{i<j=1}^{4}\sum_{a=1}^{8}(\lambda_i^a \cdot \lambda_j^a) V_{ij}(\boldsymbol{r}_{ij}),
\end{aligned}
\end{eqnarray}
where $m_i$ and $\boldsymbol{p}_i$ are the mass and momentum of the $i^{th}$ quark, respectively, 
and $T_G$ is the kinetic energy of the center-of-mass motion that must be subtracted 
for the calculation of tetra-quark masses. 
The SU(3) Gell-Mann matrices $\lambda_i^a$ are for the color space with color index $a (= 1, \dots 8)$ 
acting on the $i^{th}$ quark.
The potential $V_{ij}$ is taken from Ref.~\cite{SilvestreBrac:1996bg}, given by 
\begin{eqnarray}
\label{potential}
\begin{aligned}
	V_{ij}(\boldsymbol{r})=&-\frac{\kappa}{r}+\lambda r^p-\Lambda \\
	&+\frac{2\pi\kappa'}{3m_i m_j}\frac{\mathrm{exp}(-r^2/r_0^2)}{\pi^{3/2}r_0^3}({\bm\sigma}_i\cdot{\bm\sigma}_j),
\end{aligned}
\end{eqnarray}
with
$
r_0(m_i,m_j)=A[({2m_i m_j})/({m_i+m_j})]^{-B}.
$
The parameters are taken from Ref.~\cite{SilvestreBrac:1996bg} with some fine tuning 
as optimized for the present study.  
As discussed in the previous study, it is important to see how the above Hamiltonian 
reproduces the masses of relevant threshold mesons.  
For this we refer to Table 1 of Ref.~\cite{Meng:2020knc} which justifies 
the use of the present Hamiltonian.  

To solve the four-body problem for resonances, 
we employ the Gaussian expansion method~\cite{Hiyama:2003cu}
for which the variational method with the real scaling technique is utilized most efficiently.  
The variational wave function $\Psi_{I,JM}$ is formed for states of isospin $I$ and total spin $(J,M)$
as follows:
\begin{eqnarray}
	\Psi_{I, JM}= & {\cal A} \sum_C \xi_1^{(C)} \sum_{\gamma} B_{\gamma}^{(C)}\eta^{(C)}_I 
  \big[[\chi_{\frac{1}{2}}\chi_{\frac{1}{2}}]_s [\chi_{\frac{1}{2}}
 \chi_{\frac{1}{2}}]_{s'}  \big]_S
 \nonumber\\
	&\times  \big[[\phi^{(C)}_{n\ell}({\bm r}) \phi'^{\,(C)}_{NL}({\bm \rho})]_\Lambda
\psi^{(C)}_{\nu \lambda }(\bm{\lambda}) \big]_{L_{tot}} \bigg]_{JM} ,
\label{total wave function}
\end{eqnarray}
where 
$\xi_1$ stands for the color singlet wave function, 
$\eta$ for the isospin part of light anti-diquarks, 
$\chi$ for the spin part of each quark,
and 
$\phi$, $\phi'$, and $\psi$ for the spatial wave functions.
${\cal A}$ denotes the anti-symmetrization among the like quarks.
The index $\gamma$ in Eq.~(\ref{total wave function}) denotes collectively all the quantum numbers 
needed for the expansion, 
$\gamma \equiv \{s, s', S,n,N,\nu,\ell,L,\lambda, L_{tot}\}$.
Energy eigenvalues, $E$, and corresponding expansion coefficients, $B_{\gamma}^{(C)}$, are determined 
by diagonalizing the Hamiltonian matrix computed by the basis functions $\Psi_{I, JM}$. 

\begin{figure}[ht]
\centering
\includegraphics[height=3cm]{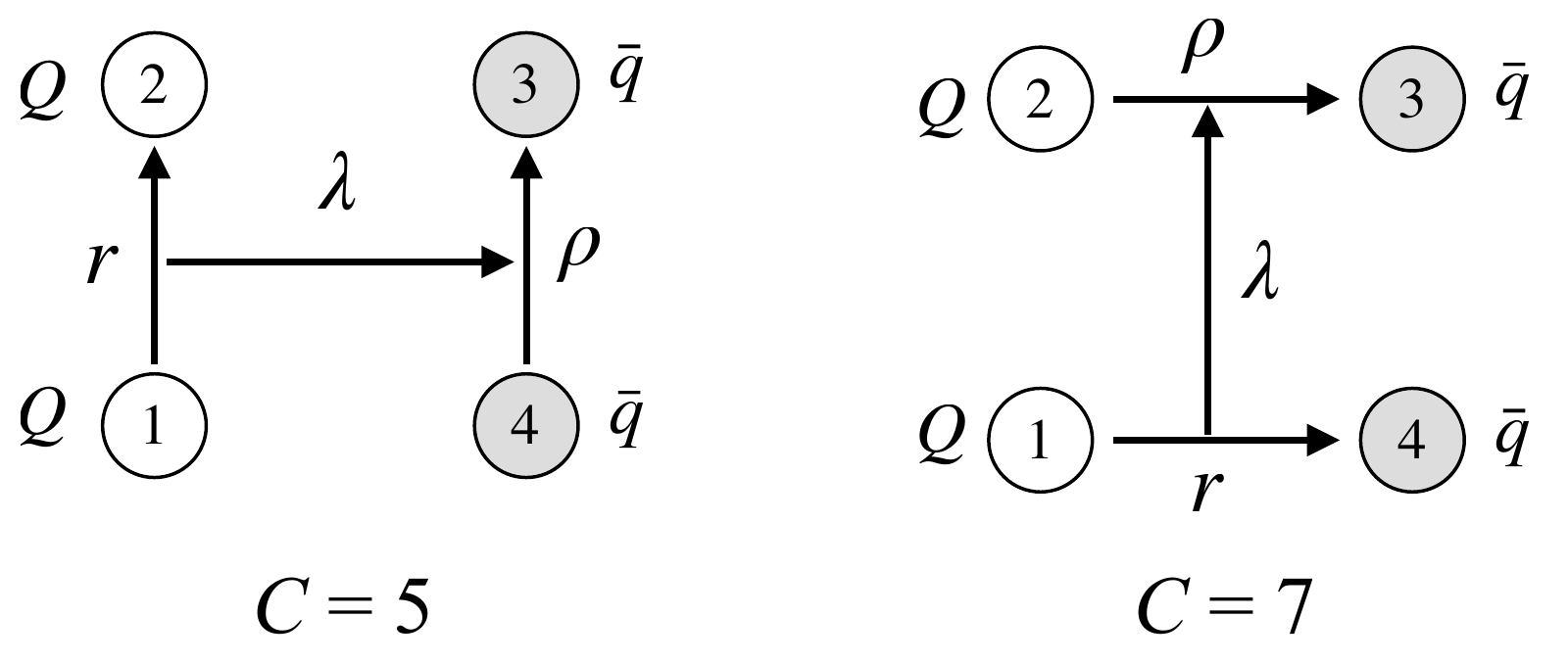}
\caption{ 
The Jacobi coordinates $C=5$ and $C=7$ for $QQ\bar q \bar q$ tetraquarks~\cite{Meng:2020knc}. 
} 
\label{fig_jacobi}
\end{figure}

The index $C (= 1, \cdots 7)$ labels the seven sets of the Jacobi coordinates
employed in the present study, 
which are illustrated in Fig.~1 of Ref.~\cite{Meng:2020knc}.
Among them, $C=5$ and 7 are shown in Fig.~\ref{fig_jacobi}.
The Jacobi coordinate $C = 7$ contains dominantly unconfined scattering 
channels, such as $BB^*$ and $B^*B^*$, with the color wave function,
$\xi_1^{C=7}=[(14)_{1}, (23)_1]_1$,
where the indices 1 and 2 are for $b(Q)$-quarks and 
3 and 4 for $\bar q$.  

The $C=7$ channel plays a key role in the scaling method 
to judge whether a state is a meson-meson scattering state or a resonant state.
The resonant states are identified by the real-scaling method, in which
the Gaussian range parameters for the $\lambda$ coordinate of $C=7$ are scaled 
as $\lambda \rightarrow \alpha\lambda$.
When $\alpha$ is varied, the distance of two color-singlet mesons in the channel $C=7$ is scaled with $\alpha$. 
Then energy eigenvalues of scattering states will decrease as $\alpha$ increases, 
while compact resonance states will stay at the same energy. 
Thus the $\alpha$ dependences of energy eigenvalues can be used 
to distinguish resonance states from scattering states. 
Furthermore, level crossing with repulsion allow us to evaluate the widths and 
the couplings of the resonance states as shown below.


\section {Results} 

\begin{figure}[ht]
\centering
\includegraphics[width=0.45\textwidth]{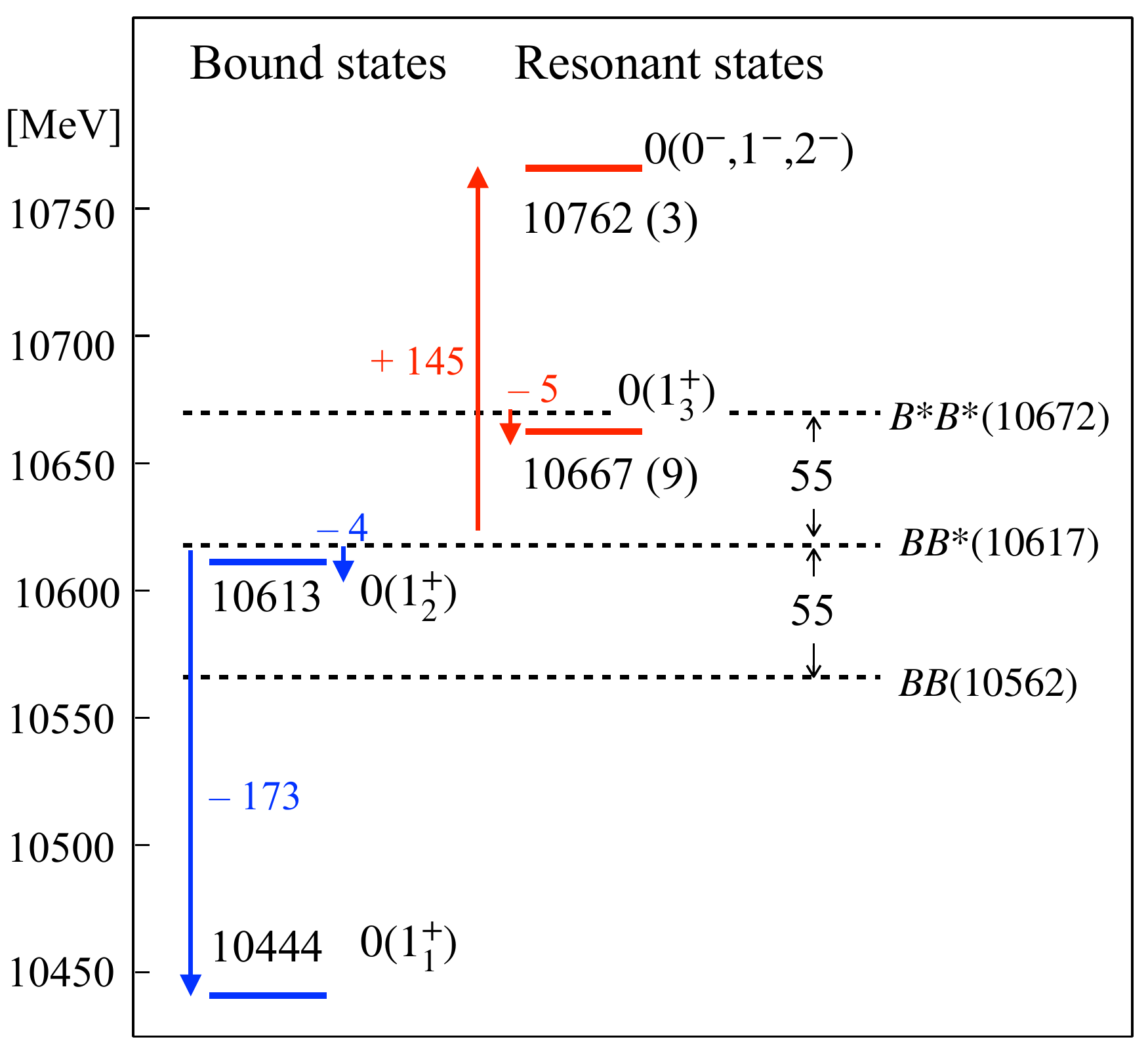}
\caption{ 
Calculated masses of $bb\bar{q}\bar{q}$ bound and resonant states.
The quantum numbers are designated as $I (J^{\pi})$.
Blue lines are bound states below the $BB^*$ threshold,  and
red lines show resonant states.
The numbers indicate absolute masses 
as well as the binding or excitation energies (in MeV), and
the numbers in parenthesis are the estimated decay widths in the
real scaling method (in MeV). 
}
\label{fig_masses}
\end{figure}

The obtained energy spectra of the resonant states together with bound states of
$bb\bar{q}\bar {q}$
are shown in Fig.~\ref{fig_masses}.
Two bound states ($1^+_1$ and $1^+_2$) with $I (J^{\pi}) =0(1^+)$ are already reported in our previous paper~\cite{Meng:2020knc}, 
a deeply bound state and a shallow bound state below the $BB^*$ threshold (blue lines).
In this study, we further find a few low-lying resonant states; 
another $I(J^{\pi})=0(1^+)$ state ($1^+_3$) located 
at 10667 MeV which is 5 MeV below the $B^*B^*$ threshold 
and negative-parity states with $I(J^{\pi})=0(0^-)$, $0(1^-)$ and $0(2^-)$, which are
located at 10762 MeV which are 90 MeV above the $B^*B^*$ threshold.

\begin{figure}[htb]
\centering
\includegraphics[width=0.45\textwidth]{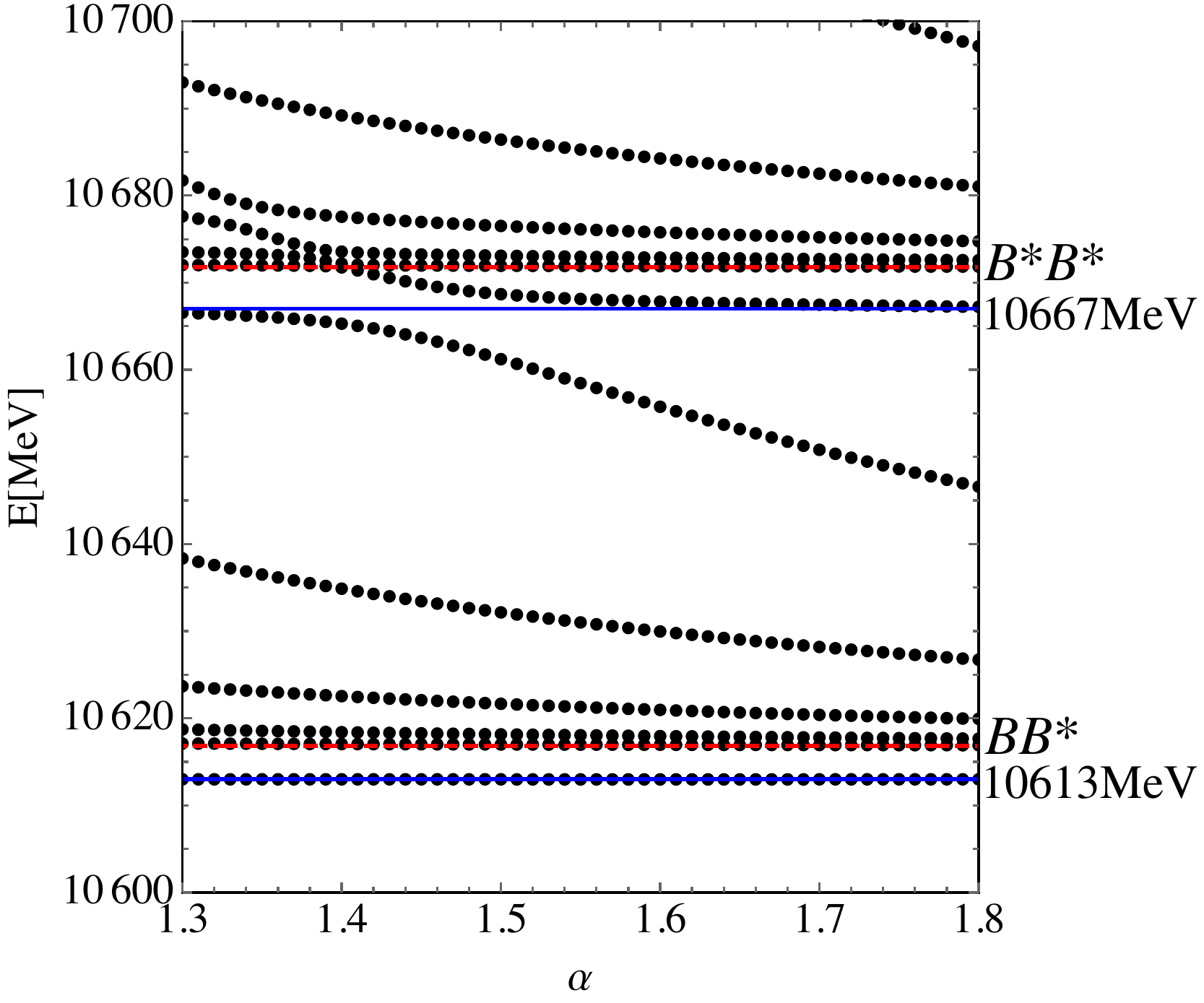}
\caption{ 
The stabilization plots of the energy eigenvalues of $I(J^{\pi})=0(1^+)$ states by changing  
the scaling parameter $\alpha$.}
\label{fig_scal}
\end{figure}
The resonant states are identified by the real-scaling method\cite{Simons:1981aa}.
In Fig.~\ref{fig_scal}, we plot the energy eigenvalues of the $1^+$ states as functions of the
scaling factor $\alpha$. 
A horizontal behavior at around 10667 MeV is identified as the $1^+_3$ resonant state.%
\footnote{Another horizontal line at 10613 MeV corresponds to the bound state $1^+_2$.}
One sees a level crossing of the resonance with a $BB^*$ continuum state at
around $\alpha=1.45$.
The width of the resonance can be roughly estimated from the two-state crossing formula given 
in Ref.~\cite{Simons:1981aa}, as
\begin{eqnarray}
\Gamma\approx 2\Delta E\times\frac{\sqrt{|S_r||S_c|}}{|S_r-S_c|} 
\label{Simon formula}
\end{eqnarray}
where $S_r$ and $S_c$ are the slopes of the two crossing levels, resonance and continuum, 
respectively. $\Delta E$ is the energy difference between the upper and lower branches at 
the level crossing point.
We then estimate the decay width to be 9 MeV for the $1^+_3$ resonant state.


Next, let us study the structure of each resonant state in detail.
First, we focus on the  $1^+_3$ state, located at 10667 MeV, 
that is 50 MeV above the lowest $BB^* (1^+)$ threshold.
This state is compared with the lower $1^+$ states, 
the lowest deeply bound state ($1^+_1$) and 
the shallow bound state just below the $BB^*$ threshold ($1^+_2$) . 
The binding energy of the latter is very small, 4 MeV, 
and almost the same as the (quasi-binding) energy of the $1^+_3$ state measured from the $B^*B^*$ threshold.

As was shown in our previous study~\cite{Meng:2020knc}, the lowest $1^+_1$ state 
is a compact bound state with the binding energy of 173 MeV.
The large binding energy is driven  
by the short-range color Coulomb force between the two $b$ quarks, which forms
the 1S state, $bb (1S)$.
In contrast, the second and third states are shallow and are expected 
to have strong effects from the near-by thresholds, $BB^*$ and $B^*B^*$.

\begin{figure}[htb]
\centering
\includegraphics[width=0.4\textwidth]{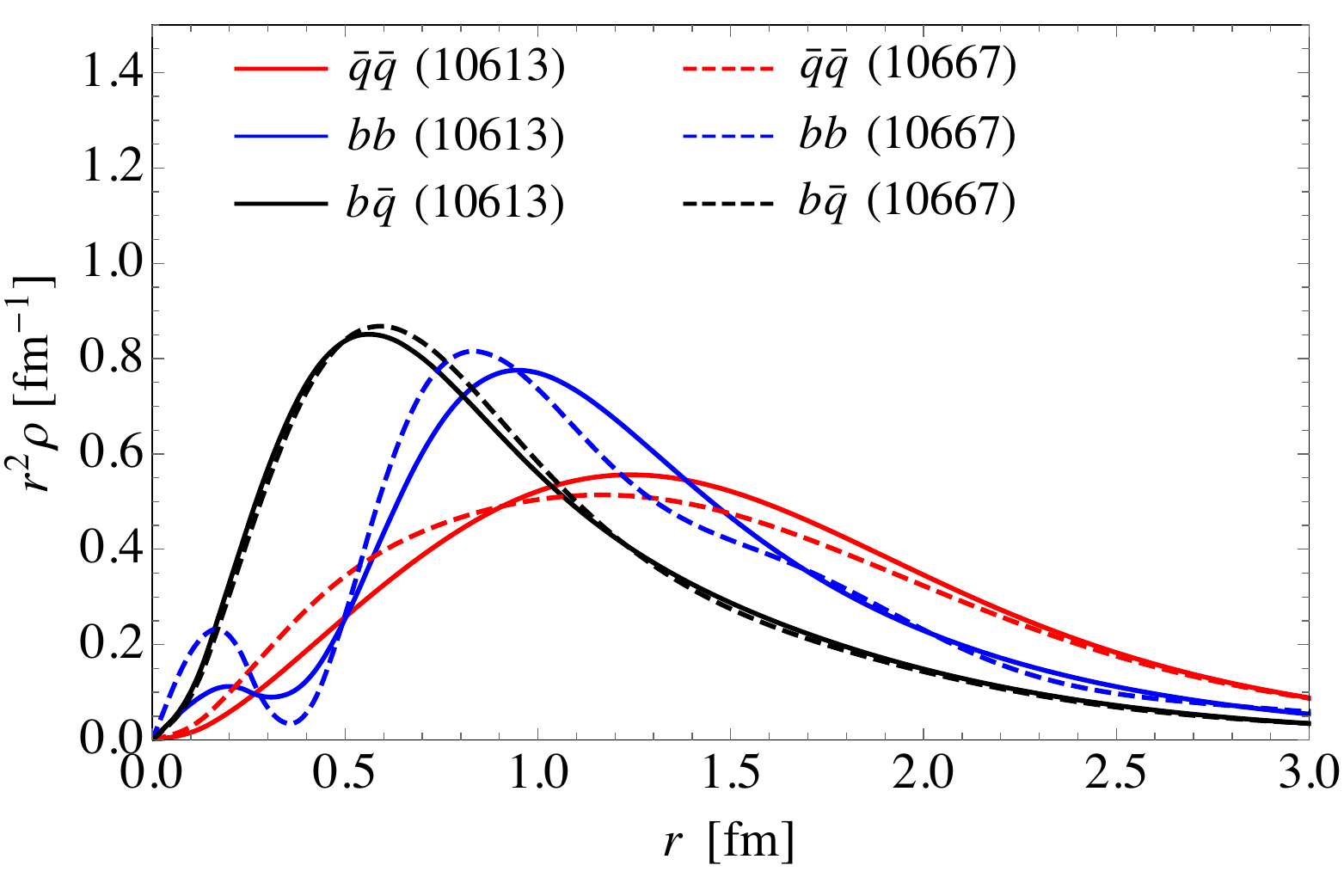}
\caption{ 
Probability densities of $bb$(blue), $b\bar{q}$(black) and 
$\bar{q}\bar{q}$(red) of $1^+_2$ (10613 MeV, solid lines) and $1^+_3$ (10667 MeV, dashed lines) states. 
} 
\label{fig_den1p}
\end{figure}

In order to reveal their dynamical structure, we 
calculate the two-body density distributions.
Fig.~\ref{fig_den1p} compares the distributions as functions of distances between designated quark pairs
in the $1^+_2$ and $1^+_3$ states.
One sees long tail regions in the $bb$ and $\bar q\bar q$ distributions, that look almost identical 
at $r$ larger than 1.5 fm for the $1^+_2$ and $1^+_3$ states.
It is clear that due to color confinement the tail parts are dominated 
by the color-singlet component, that is, $BB^*$ and $B^*B^*$ in these cases.
Calculating the overlap probabilities with the $S$-wave $BB^*$, $P(BB^*)$ and $B^*B^*$ states, $P(B^*B^*)$,%
\footnote{It should be noted here that the probabilities for the $1^+_3$ resonant state are 
defined for the discrete state, whose wave function is
truncated at around 2.4 fm due to the the choice of the basis function of the Gaussian expansion.}
we find that the ratio $P(B^*B^*)/P(BB^*)$ is about 0.02 for $1^+_2$ and $\sim 2.4$ for $1^+_3$.
So the tail part of the $1^+_2$ state is dominated by $BB^*$, while that of $1^+_3$ is by $B^*B^*$.

For further analyses of the inner parts, 
we compute expectation values of the color operator, $\langle\lambda_{b1}\cdot\lambda_{b2}\rangle$, of the $b_1b_2$ sub-system.
The results are $\sim 0$ for the $1^+_2$ state and $-0.63$ for $1^+_3$.
Note that the tail part, consisting of $BB^*$ and $B^*B^*$ molecule components, 
should give $\langle\lambda_{b1}\cdot\lambda_{b2}\rangle=0$.
Therefore the vanishing expectation value indicates that 
the $1^+_2$ state is almost purely $BB^*$ molecular bound state.

On the other hand, the large negative value of $\langle\lambda_{b1}\cdot\lambda_{b2}\rangle$
indicates that the $1^+_3$ state contains
a significant color $\bar{\bm 3}-{\bm 3}$ tetraquark component at short distances.
As the $1^+_3$ state should be orthogonal to the compact  
$1^+_1$ state, the tetraquark component of $1^+_3$ is supposed to be
the first excited state, that is $bb(2S)$ combined with the ground $\bar q\bar q$ state.
This picture is consistent with the density distribution given in Fig.~\ref{fig_den1p}, which 
shows a nodal behavior in the $bb$ distribution at small $r$.
The large $P(B^*B^*)/P(BB^*)$ ratio and the small width indicate that 
the tail part of the $1^+_3$ state is dominated by the $B^*B^*$ molecular component.%
\footnote{The $1^+_3$ state contains also a significant $BB^*$ component (about 30 \% for the truncated
wave function),
as it eventually falls apart into the $S$-wave $BB^*$ channel.}
We thus conclude that the $1^+_3$ state is an admixture of the compact $bb(2S)(\bar q\bar q)$ tetraquark 
and an extended $B^*B^*$ molecular component.

Next, we consider the structure of the lowest negative-parity states 
with $J^{\pi}=0^-,1^-,2^-$ 
located at 10762 MeV which is 145 MeV above the $BB^*$ threshold.
In this calculation, we do not introduce spin-orbit interaction and 
thus the $0^-$, $1^-$ and $2^-$ states are degenerate.%
\footnote{The decay widths, $\sim 3$ MeV, of these triplet states are also identical, 
because, without spin-orbit interactions,
the $0^-$, $1^-$ and $2^-$ states decay only to the $BB^* (I=0, L=1)$ channel with the same decay rates.
}
The main configuration of these states is given by the Jacobi coordinate $C=5$ (See Fig.~\ref{fig_jacobi}).
The $P$-wave excitation is located in the $\lambda$ coordinate 
between the $bb$ and $\bar q\bar q$, and thus this is called $\lambda$-mode
excited state.
The main component is given by the following configuration; 
the relative angular momentum $\ell$ for $bb$ is 0,
$L$ for $\bar{q}\bar{q}$ is 0, and $\lambda$ for $bb-\bar{q}\bar{q}$ is 1 
in Eq.~(\ref{total wave function}).

It should be emphasized that these $\lambda$-mode excitations form 
a heavy-quark spin triplet ($0^-$, $1^-$ and $2^-$)
under the heavy-quark spin symmetry, which is understood as follows:
For $P$-wave excitations in the $\lambda$-mode, spin of the light anti-diquark is $0$, which is combined 
with the orbital angular momentum to make spin of the light cloud surrounding the heavy diquark be $1$. 
This is combined with the spin $1$ of $bb$ diquark to form tetraquark states of $0^-$, $1^-$ and $2^-$ as a heavy-qurak spin triplet.
Although they are exactly degenerate in the present analysis, the degeneracy 
could be resolved in the reality due to spin-orbit interaction.
Still we expect that mass difference will be tiny since the difference is suppressed by $1/m_b$ due to the heavy quark symmetry.

\begin{figure}[ht]
\centering
\includegraphics[width=0.4\textwidth]{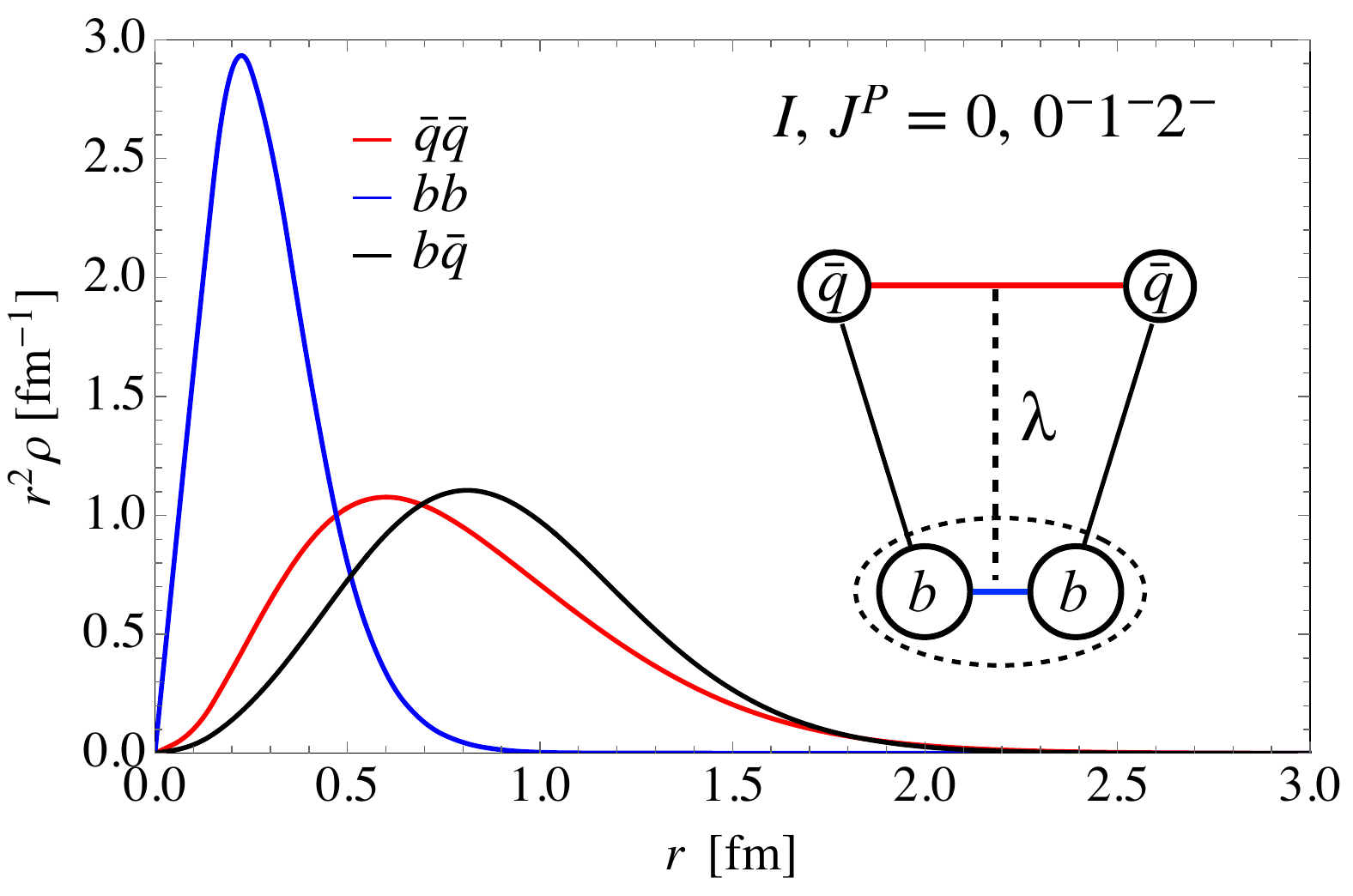}
\caption{ 
Density distributions of $bb$(blue), $b\bar{q}$(black) and
$\bar{q}\bar{q}$(red) of the lowest negative parity resonant state.
A sketch is drawn for the $\lambda$-mode like configuration. 
} 
\label{fig_den1m}
\end{figure}

Fig.~\ref{fig_den1m} shows the density distributions of this state.
We see that the $bb$ density is compact and that of $bb-\bar{q}\bar{q}$
is dilute showing the relative $P$-wave behavior.
By these density distributions, we conjectures that this $\lambda$-mode state has a three-body-like structure, 
a $bb$ diquark plus two light anti-quarks, illustrated as an inset of Fig.~\ref{fig_den1m}.
This picture is consistent with the small decay width to the $BB/BB^*/B^*B^*$ channels.
As the two-body decays are in relative $P$-wave states in which two $b$ quarks are separated,
the overlap to the three-body-like resonant state with a compact $bb$ wave function must be small.




It should be stressed that these narrow three-body-like states are located much lower 
than the other negative-parity states.
In the present study we have not discussed other possible excitations corresponding 
to $\rho$-mode excitations, which are internal excitations of light anti-diquark.  
In numerical analysis, we have observed some signals for the $\rho$-mode excitations 
at higher energies at around and over 11000  MeV, which is about 300 MeV higher 
than the $\lambda$-mode heavy-quark triplet.  
However, these states will likely be affected by possible pion emission decays, 
which is beyond the present scope and therefore will be discussed elsewhere. 


\section*{Acknowledgments}

This works is supported in part by Grants-in Aid for Scientific Research on Innovative Areas, No. 18H05407 for QM, EH, AH, No. 20K03927 for MH, and Nos. 19H05159, 20K03959 and 21H00132 for MO.  

\bibliography{references}

\end{document}